\documentclass[letterpaper, 10 pt, conference]{ieeeconf}  

\IEEEoverridecommandlockouts                              

\usepackage{amsmath} 

\usepackage{graphicx}
\usepackage{subcaption} 
\usepackage{hyperref}
\usepackage{multirow}
\usepackage{booktabs}
    
\title{\LARGE \bf
TR-LLM: Integrating Trajectory Data for Scene-Aware LLM-Based Human Action Prediction
}

\author{Kojiro Takeyama$^{1,2}$ Yimeng Liu$^{1}$ and Misha Sra$^{1}$
\thanks{$^{1}$Kojiro Takeyama, Yimeng Liu, and Misha Sra are with the Department of Computer Science, University of California Santa Barbara,
        Santa Barbara,  CA 93106-5080, USA
        {\tt\small takeyama@ucsb.edu, yimengliu@ucsb.edu, sra@ucsb.edu}}%
\thanks{$^{2}$Kojiro Takeyama is with Toyota Motor North America, Ann Arbor, MI 48105-9748, USA
        {\tt\small kojiro.takeyama@toyota.com}}%
}

\begin{document}

\maketitle
\thispagestyle{empty}
\pagestyle{empty}

\begin{abstract}

Accurate prediction of human behavior is crucial for AI systems to effectively support real-world applications, such as autonomous robots anticipating and assisting with human tasks. Real-world scenarios frequently present challenges such as occlusions and incomplete scene observations, which can compromise predictive accuracy. Thus, traditional video-based methods often struggle due to limited temporal and spatial perspectives. Large Language Models (LLMs) offer a promising alternative. Having been trained on a large text corpus describing human behaviors, LLMs likely encode plausible sequences of human actions in a home environment. However, LLMs, trained primarily on text data, lack inherent spatial awareness and real-time environmental perception. They struggle with understanding physical constraints and spatial geometry. Therefore, to be effective in a real-world spatial scenario, we propose a multimodal prediction framework that enhances LLM-based action prediction by integrating physical constraints derived from human trajectories. Our experiments demonstrate that combining LLM predictions with trajectory data significantly improves overall prediction performance. This enhancement is particularly notable in situations where the LLM receives limited scene information, highlighting the complementary nature of linguistic knowledge and physical constraints in understanding and anticipating human behavior.

Project page: 

\url{https://sites.google.com/view/trllm?usp=sharing}

Github repo:

\url{https://github.com/kojirotakeyama/TR-LLM/blob/main/readme.md}
\end{abstract}

\section{INTRODUCTION}

Predicting human behavior is essential for AI systems to integrate seamlessly into our lives and provide effective support. This capability is particularly crucial for applications like support robots, which need to anticipate human actions and proactively perform tasks in a home environment. While traditional approaches have relied heavily on video data for action prediction \cite{c6,c16}, they often fall short due to their limited temporal and spatial scope. Human behavior is influenced by complex environmental and personal factors that extend beyond what can be captured in video footage. These include temporal elements like time of day, environmental conditions such as room temperature, the state and position of objects in the scene, and human-specific attributes including personal characteristics, action history, and the relationship with others present in the room. These factors exert significant influence on human behavior as exemplified by the contrast between retrieving an item from a fridge during a time-constrained morning routine versus during a more leisurely evening meal preparation or the contrast between weekday vs weekend routines. To truly understand and predict human behavior, we must take into account this broader contextual landscape, moving beyond the constraints of conventional video-based methods.

\begin{figure}[!t]
  \includegraphics[width=\columnwidth]{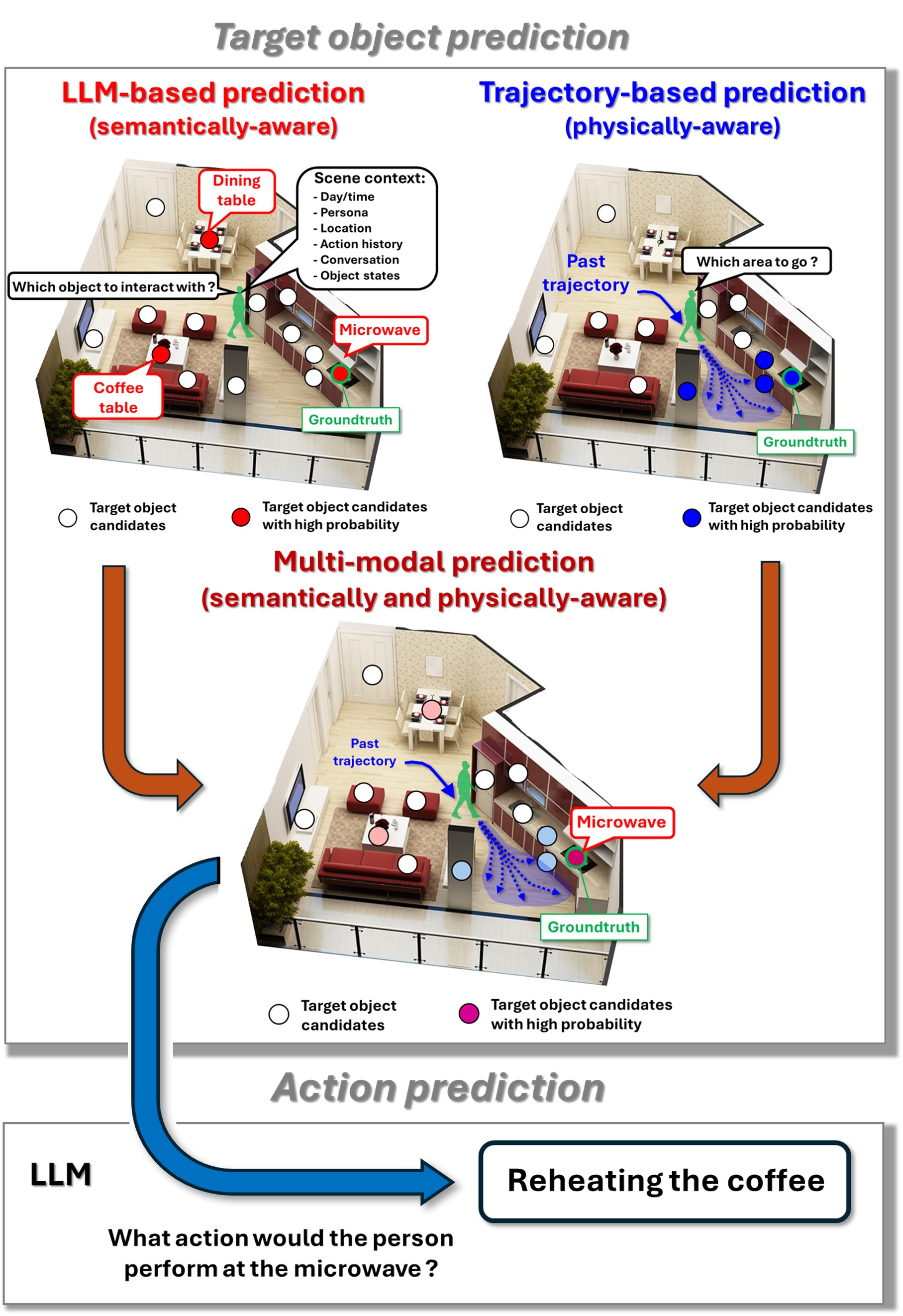}
  \caption{Overview of our approach: We propose a multi-modal action prediction framework that incorporates both an LLM and human trajectories. The core idea is to integrate two different perspectives—physical and semantic factors—through an object-based action prediction framework to reduce uncertainties and enhance action prediction accuracy.}
  \label{fig:overview}
\end{figure}


Recent studies have explored the use of Large Language Models (LLMs) for predicting human actions \cite{c20,c21}. These models leverage extensive knowledge of human behavior in home environments to forecast actions across diverse scene contexts and generally perform robustly even in zero-shot settings. 
However, observable scene contexts in the real world are often insufficient for predicting a person's intentions accurately. For instance, scene context derived from sensor data (e.g., cameras, microphones) frequently suffers from limitations in coverage, sensitivity, or resolution, leading to increased uncertainty in predictions. This presents an inherent limitation in LLM-based action predictions.

To address this shortcoming, we propose a multi-modal prediction framework that enhances LLM-based action prediction by integrating physical constraints. Our approach uses a person's past trajectory to infer their likely destination, thereby imposing physical constraints on their next target object and narrowing down potential actions (Fig.\ref{fig:overview}). Recognizing that target area prediction is highly dependent on factors such as room layout, location, and speed, we leverage an indoor human locomotion dataset \cite{c33} to learn these complex relationships, enabling us to derive a reliable probabilistic distribution of target areas.

Unlike vision-language models (VLMs) \cite{c34}, which are adept at recognizing and interpreting images and text yet often struggle to predict physical quantities under complex image constraints, our method incorporates a dedicated physically-aware prediction module trained on practical datasets to overcome these challenges.

The contribution of our work is described as follows.

\begin{itemize}
    \item Input: Incorporating LLMs to human action anticipation considering a wide variety of semantic scene context

    \item Proposing a multi-modal action prediction framework incorporating LLMs with trajectory that imposes physical constraints on the person's next target object, helping to narrow down potential actions.

    \item Building an evaluation dataset for our multi-modal approach, which includes scene map, trajectory, and scene contexts.

    \item Demonstrating that our method markedly enhances prediction performance compared with LLM and VLM, especially in scenarios where the LLM has limited access to scene information. 
\end{itemize}

\section{Related works}
\subsection{Vision-based approach}
Previous works have primarily focused on vision-based human action prediction. One line of research has focused on video-based action prediction, utilizing the recent past sequence of video frames to forecast future actions. Given the availability of large-scale action video datasets (e.g., Ego4D \cite{c2}, Human3.6M \cite{c9}, Home Action Genome \cite{c1}), many researchers have adopted machine-learning-based approaches \cite{c8,c7,c5,c4}, often employing sequential models such as LSTM \cite{c6} or transformers \cite{c15,c14}.
Some studies have integrated multimodal models to improve performance, combining video with other modalities (e.g., video-acoustic models \cite{c3}, video-text models \cite{c11,c10}). Others have focused on object-based cues within the video to predict future actions \cite{c12,c13}.
Another research direction involves predicting future motions from past motions, using only the human body's pose data. For instance, \cite{c19} introduced a stochastic model for predicting diverse future actions with probability, while \cite{c17,c18} incorporated graph convolutional networks with attention mechanisms to improve prediction accuracy. Additionally, \cite{c16} proposed the use of a diffusion model to account for multi-person interactions.

In existing video-based action prediction research, a limitation lies in the spatial and temporal restrictions of scene context. Most studies focus on information within the limited field of view of ego-centric or third-person cameras, which only provides localized data. This narrow perspective hinders understanding of broader, more complex behaviors in real-world environments.
Additionally, privacy concerns restrict the observation of individuals' daily activities over extended periods in uncontrolled settings, limiting the ability to learn a diverse range of daily action patterns from existing datasets. To overcome these constraints, we propose leveraging large language models (LLMs), which encode vast everyday knowledge in natural language. By incorporating LLMs, we remove spatial and temporal limitations, enabling a more flexible and generalized approach to action prediction that captures broader patterns applicable to diverse real-world scenarios.

\subsection{Language model-based approach}
With the rise of large language models (LLMs), many studies have leveraged LLMs to analyze social human behaviors. Notably, \cite{c23, c22} made a significant impact on the field by simulating town-scale, multi-agent human daily life, incorporating social interactions between agents throughout the entire pipeline using LLMs. This work has inspired subsequent efforts to extend the scale and complexity of such simulations \cite{c24, c25, c26}.
However, a notable gap exists between their work and ours, as their focus is primarily on macroscopic-scale social behavior simulations. Additionally, their approach centers on intention-conditioned action generation within these simulations, which contrasts with our goal of predicting human actions in real-world scenarios.

Recently, several studies have explored LLM-based human action prediction. Specifically, \cite{c20, c21, c35} used LLMs to predict human actions, enabling robots to plan supportive responses. In these approaches, scene context was provided as textual input to the LLM for predicting subsequent human actions. However, while these studies incorporated contextual information, they neglected physical constraints such as realistic human motion dynamics or spatial environment structures. Meanwhile, \cite{c39} addressed human motion by introducing a graph-based transition model, but their representation of human movements was heuristic and simplistic, lacking grounding in real-world human trajectory data.

Furthermore, vision-language models (VLMs)\cite{c34} have recently emerged as a prominent technology. These models possess the capability to recognize and understand both images and text, and they are utilized in a wide range of applications such as action prediction\cite{c38} or spatial reasoning\cite{c36,c37}. Although they could potentially be applied to our task, pre-trained VLMs generally lack the ability to predict physical quantities under complex image constraints since they are primarily trained to establish correspondences between images and text. In contrast, our method can evaluate complex image constraints and predict the target object by employing a physically-aware prediction module that operates independently of the LLM module.

\section{Method}
\label{sec:method}
\subsection{Problem definition}
\begin{figure}
  \includegraphics[width=\columnwidth]{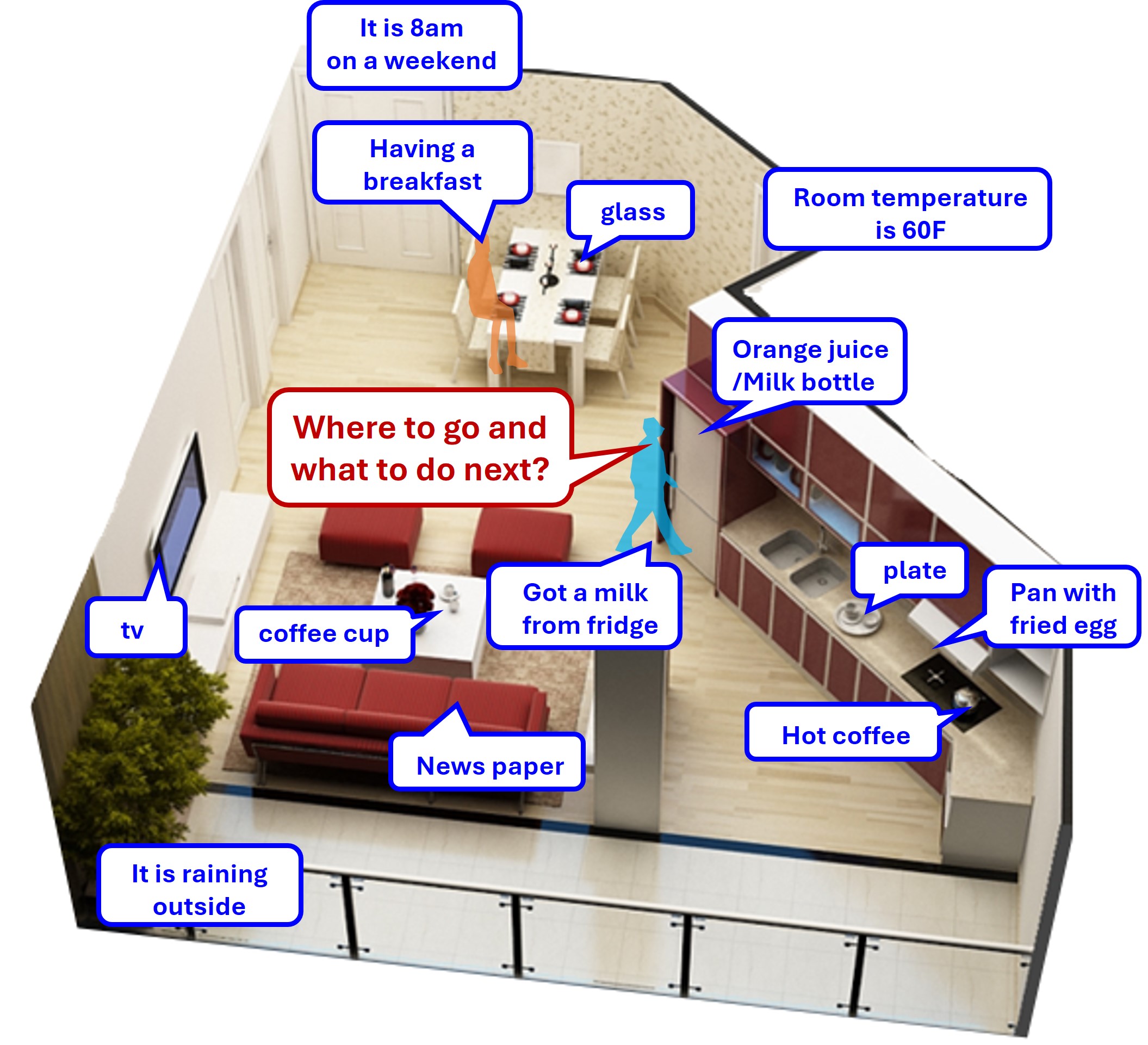}
  \caption{Scene states in a room}
  \label{fig:problem definition}
\end{figure}

In daily life, individuals perform sequential actions while interacting with various objects. For instance, one might take a glass from a cupboard, fill it with a drink from the fridge, and then sit on the couch to drink. In this study, we focus on predicting the transition between these sequential actions, specifically forecasting the next action as a person moves toward the target object.
We assume the scene is a room-scale home environment where we can observe various scene states except for the persons' internal states (Fig.\ref{fig:problem definition}).
The input comprises both text-based and image-based scene contexts. The text-based context includes semantic details—such as the time of day, object lists, a person's action history, and conversations—while the image-based context provides physical scene information through a 2D image that shows an object layout map and the person's trajectory. In our task, we first predict the target object using both contexts and then forecast the future action based on this prediction.

\subsection{Method overview}
Fig.\ref{fig:overview} illustrates an overview of our approach. We propose a multi-modal action prediction framework that incorporates both an LLM and human trajectories. The core idea is to integrate two different perspectives—physical and semantic factors—through an object-based action prediction framework.
Our framework consists of two primary steps: target object prediction (\ref{sec:target object prediciton}) and action prediction (\ref{sec:action prediction}).
In the target object prediction step, we first utilize a LLM to predict a person's target object based on the input scene context, generating a probability distribution over potential objects in the room from a semantic perspective (\ref{sec:llm-based prediciton}). Subsequently, we incorporate the person's past trajectory to infer their likely destination, applying physical constraints to refine the prediction of the next target object (\ref{sec:trajectory-based prediciton}).

In the action prediction phase, the individual's subsequent action is predicted using a LLM, based on the identified target object.

Further details of the approach are provided in the following sections.

\subsection{Target object prediction}
\label{sec:target object prediciton}
\subsubsection{LLM-based prediction}
\label{sec:llm-based prediciton}
\begin{figure}
  \includegraphics[width=\columnwidth]{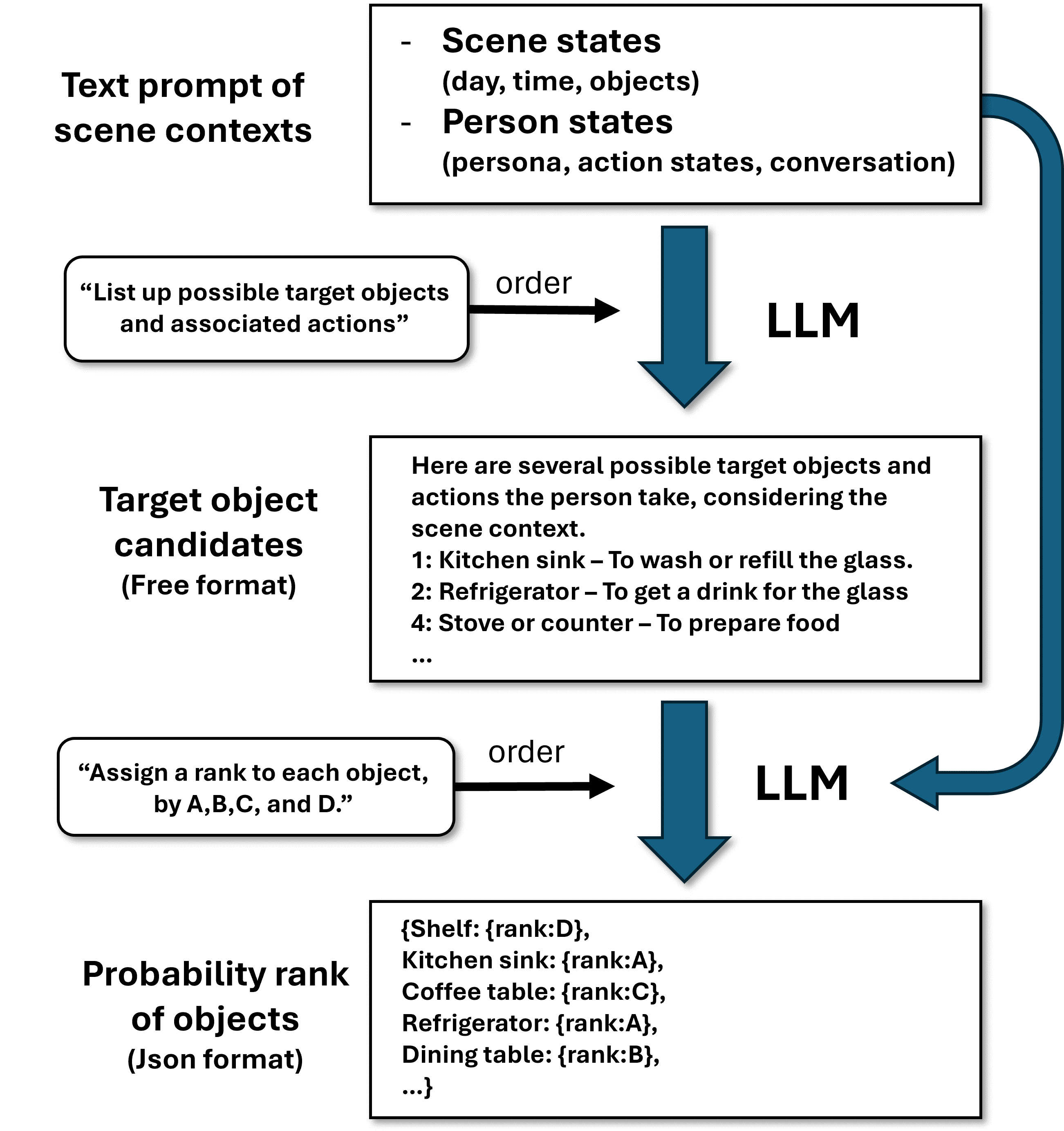}
  \caption{Pipeline of LLM-based prediction}
  \label{fig:LLM pipeline}
\end{figure}

Fig.\ref{fig:LLM pipeline} shows the pipeline of LLM-based target object prediction. In the initial step, various scene contexts—such as time of day, the person's state (including location, action history, and conversation), and a list of objects in the room—are provided as a text prompt and input into the LLM. Additionally, an order prompt is used to guide the LLM in generating a comprehensive, free-form list of potential target objects and their associated actions. This intermediate step facilitates the identification of probable target objects by considering their corresponding actions. Consequently, we derive a ranking of objects in the environment based on the target object candidates given in the previous step and the initial scene context prompt. The ranking is divided into four probability levels: A (high probability), B (moderate probability), C (low probability), and D (very low probability). The output from the LLM is subsequently converted into numerical scores and normalized into probabilities that sum to 1. In this study, scores of 15, 10, 5, and 1 were assigned to categories A, B, C, and D, respectively.

\subsubsection{Trajectory-based prediction}
\label{sec:trajectory-based prediciton}
We utilize past trajectory data and scene map information to constrain the target area of a person. Since the target area prediction is heavily influenced by factors such as room layout, the person's location, and speed, we leverage the LocoVR dataset\cite{c33} to learn these intricate relationships. The LocoVR dataset consists of over 7,000 human trajectories captured in 131 indoor environments, with each trajectory segment spanning from a randomly assigned start to a goal location within the room. We employed a simple U-Net model to identify potential target areas based on past trajectories and scene maps. Specifically, the model takes as input a binary obstacle scene map (HxWx1) and a trajectory represented as points on time-series images (HxWxt), and it outputs the target object's position as an image (HxWx1). During training, we compute the binary cross entropy loss between the predicted and ground truth target object positions represented as the image.
The trained model predicts the probabilistic distribution of the target area, and we calculate the overlap between this predicted distribution and the spatial location of objects in the scene to estimate the probability of each object being the target.

\begin{figure}
  \includegraphics[width=\columnwidth]{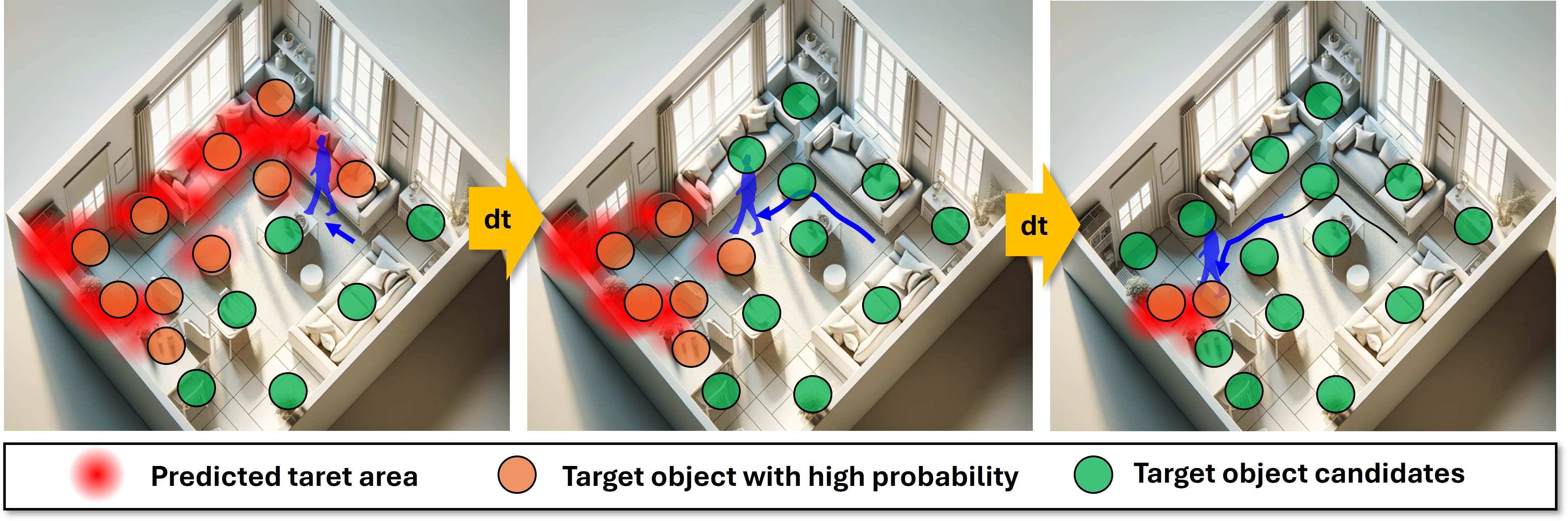}
  \caption{Trajectory-based target object prediction: At the beginning of the trajectory (left image), the potential target area is widely distributed. As the person advances, both the predicted target area and the range of potential target objects become progressively more focused (center and right images).}
  \label{fig:Goal prediction}
\end{figure}

Fig.\ref{fig:Goal prediction} illustrates the process of trajectory-based target object prediction. In an early stage of trajectory (left image), the potential target area is broadly distributed. As the person progresses, the predicted target area and potential target objects are progressively narrowed down (center and right images). This refinement is based on a learned policy reflecting typical human behavior: when moving toward a specific goal, individuals generally avoid retracing their steps or taking unnecessary detours.

Since both LLM-based prediction and trajectory-based prediction assign probabilities to each object, we integrate these by multiplying their respective probabilities. This approach allows us to refine and identify target object candidates that are highly probable from both semantic and physical perspectives.

\subsection{Action prediction}
\label{sec:action prediction}
At this stage, we employ the LLM to predict the action corresponding to the target object identified in the previous step. The input to the LLM includes the scene context and the predicted target object, and the LLM outputs the most plausible action the person is likely to perform within the given scene context.

\section{Experiment}
\subsection{Evaluation data}

To evaluate our multimodal method, we required input data with a text prompt describing the scene context, trajectory, and scene map. As no existing dataset met these criteria, we constructed a new dataset for evaluation.

\begin{figure}
  \includegraphics[width=\columnwidth]{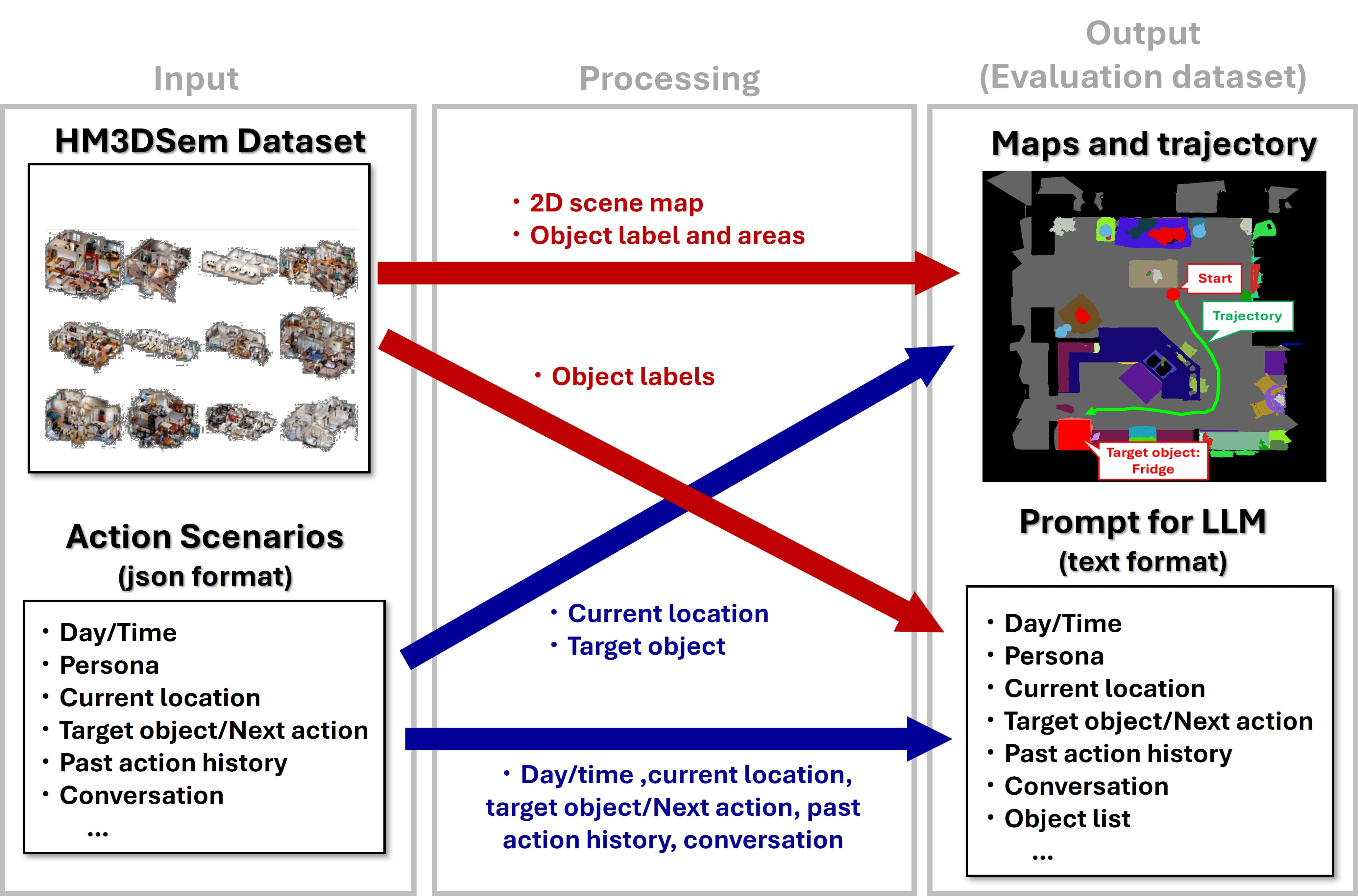}
  \caption{Pipeline of evaluation data generation}
  \label{fig:data generation pipeline}
\end{figure}

Fig.\ref{fig:data generation pipeline} illustrates the pipeline we employed to generate the evalution dataset. We utilized the Habitat-Matterport 3D Semantics Dataset (HM3DSem) \cite{c27} to obtain scene map and object data in home environments. HM3DSem provides a diverse set of 3D models of home environments, with semantic labels. From HM3DSem, we extracted scene maps represented as binary grids, where a value of 1 indicates areas walkable areas (within 0.3 meters of the floor height), and 0 corresponds to all other regions. These maps have dimensions of 256 x 256 pixels, mapping to a physical space of 10m x 10m. Additionally, we identified the pixel regions corresponding to each object within the scene map.

Furthermore, we manually designed daily action scenarios, detailing factors such as day/time, persona, current location, target object, subsequent actions, conversations, and more. These scenarios were transformed into text prompts for input to the LLM. 
We generated a trajectory from the current location to the target object by leveraging a model trained on the LocoVR dataset, which incorporates the scene map to guide the path (visit \href{https://sites.google.com/view/tr-llm}{our website}\cite{c32} for the detail).

Finally, we obtained 67 data pairs from the combination of 10 action scenarios and 9 scene maps. The total number of pairs is lower than the combination count because cases where objects from the action scenarios were absent in the scene maps were excluded.

\subsection{Benchmarks}
Since action prediction using room-scale scene context is a novel task, few existing studies could directly apply to our work. Therefore, we introduce the following three baseline methods as benchmarks:

\textbf{LLM:} In this method, the LLM predicts the target object and its corresponding human action based on a text-based scene context. The input comprises day/time information, action history, conversation, and an object candidate list. The LLM produces scores for each target object and selects the predicted action associated with the highest-scoring target object. This approach is equivalent to the LLM-based action prediction modules described in \cite{c20,c21}, though our implementation differs because the original codes were unavailable.

\textbf{VLM:} We employed pre-trained VLM models to benchmark baseline performance on multimodal data. The VLM input consists of a text-based scene context and an image that displays the scene map and the trajectory. In the image, each object and trajectory is uniquely color-coded, and a color list of the objects (e.g., sink: cyan, table: gold, trajectory: red) is included in the text-based scene context to help the model accurately identify object locations and trajectories. The VLM then outputs scores for the target objects and selects the corresponding action for the most plausible object, similar to an LLM.

\textbf{Trajectory:} We evaluated a standalone trajectory-based method, which is a component of our approach. This serves as an ablation study to assess how our method improves performance on the target object prediction compared with standalone trajectory-based method.

For both LLM and VLM, we used two types of pre-trained models—ChatGPT-4o-mini and Llama3.2—to evaluate performance across different model architectures.

\subsection{Experimental setup}
We provided three types of text-based scene context as input data to assess the model's robustness under degraded conditions: (1) full scene context, (2) scene context without conversation, and (3) scene context without conversation and past action history. These variations simulate real-world scenarios where certain information may be unavailable. The evaluation examines how our method mitigates performance degradation, demonstrating its robustness to noise in practical settings.
Additionally, we evaluate performance under three conditions of trajectory progress (d): $d>1\,\text{m}$, $d>2\,\text{m}$, and $d>3\,\text{m}$. The performance differences across these conditions reveal how the trajectory information contributes in each method.

\subsection{Evaluation tasks}
We conducted evaluations on both target object prediction and action prediction tasks. For both tasks, we conducted three trials for each evaluation and used the average of these trials as the final result to ensure the reproducibility.

\textbf{Target object prediction:}
This task serves as an essential intermediate module for action prediction. The objective is to identify the ground truth target object from a set of 30–40 objects in each scene. For evaluation, we used the top-5 accuracy metric, which considers a prediction correct if the ground truth object appears among the top five predicted objects.

\textbf{Action prediction:}
In this task, we evaluate action prediction accuracy by computing the cosine similarity between the predicted actions and their corresponding ground truth values using the SBERT library.

\subsection{Results}
\subsubsection{Quantitative result}
Table\ref{tab:accuracy_tobj} presents the accuracy of target object prediction. We describe the following conclusions based on the table\ref{tab:accuracy_tobj}:

(a) Our method outperforms both LLM and VLM across GPT and Llama models. Notably, as the available text-based scene context decreases, the prediction accuracy of both LLM and VLM deteriorates significantly, whereas our method maintains robust performance. This demonstrates that the physical cues extracted from trajectories effectively compensate for the reduced semantic context.

(b) VLM performance is comparable to—or slightly better than—that of LLM, and both degrade similarly as the available scene context decreases. These results suggest that pre-trained VLMs may struggle with solving physical problems under complex geometric constraints, which could limit their added value when combined with text-based scene contexts. This is because pre-trained VLMs are not designed to predict physical quantities under complex image constraints, as they are primarily trained to establish correspondences between images and text.

(c) Notably, our method's performance significantly exceeds the average of the standalone LLM and Trajectory approaches, indicating that integrating LLM and trajectory information effectively compensates for individual weaknesses while leveraging their respective strengths.

(d) Regarding trajectory progress distance (d), performance improves as d increases because longer trajectories offer more spatial cues to narrow down potential destination areas as shown in Fig\ref{fig:Goal prediction}. This trend is evident in both our method and the standalone trajectory approach. In contrast, VLM does not exhibit a clear correlation between traveled distance and accuracy, suggesting that it struggles to process physical information effectively.

Table\ref{tab:accuracy_act} displays the performance on action prediction. A similar trend to that observed in target object prediction is evident, emphasizing that the target object is a key cue for forecasting future actions. These results underscore the effectiveness of our approach compared to other baselines.
\newline



\begin{table}[!ht]
    \centering
    \caption{Accuracy of target object prediction}
    \scalebox{1.0} {
    \begin{tabular}{cccc}
        \toprule
        \multirow{2.5}{*}{$Method$} & \multicolumn{3}{c}{$Accuracy [\%]$} \\ \cmidrule(lr){2-4}
        & {All} & {wo/conv} & {wo/conv-hist} \\ \midrule
         LLM-llama  & 43.2 & 35.0 & 26.5 \\ \midrule
         VLM-llama ({$d>1m$})  & 40.3 & 31.6 & 31.4 \\ \midrule
         VLM-llama ({$d>2m$})  & 40.9 & 32.2 & 31.2 \\ \midrule
         VLM-llama ({$d>3m$})  & 41.4 & 32.0 & 31.2 \\ \midrule
         Ours ({$d>1m$})\\(LLM-llama+Trajectory) & 47.9 & 45.9 & 41.4 \\ \midrule
         Ours ({$d>2m$})\\(LLM-llama+Trajectory) & 49.9 & 48.1 & 43.5 \\ \midrule
         Ours ({$d>3m$})\\(LLM-llama+Trajectory) & \textbf{51.4} & \textbf{49.8} & \textbf{46.2} \\ \midrule \midrule
         LLM-gpt  & 44.7 & 37.2 & 28.0 \\ \midrule
         VLM-gpt ({$d>1m$})  & 50.9 & 40.9 & 27.8 \\ \midrule
         VLM-gpt ({$d>2m$})  & 51.5 & 42.4 & 30.3 \\ \midrule
         VLM-gpt ({$d>3m$})  & 51.6 & 43.0 & 31.0 \\ \midrule
         Ours ({$d>1m$})\\(LLM-gpt+Trajectory) & 54.6 & 51.2 & 44.3 \\ \midrule
         Ours ({$d>2m$})\\(LLM-gpt+Trajectory) & 56.4 & 53.5 & 47.3 \\ \midrule
         Ours ({$d>3m$})\\(LLM-gpt+Trajectory) & \textbf{58.6} & \textbf{56.3} & \textbf{49.1} \\ \midrule \midrule
         Trajectory ({$d>1m$}) & 33.3  & 33.3 & 33.3 \\ \midrule
         Trajectory ({$d>2m$})  & 36.7 & 36.7 & 36.7 \\ \midrule
         Trajectory ({$d>3m$})  & 39.7 & 39.7 & 39.7 \\
        \bottomrule
    \end{tabular}
    }
    \label{tab:accuracy_tobj}
\end{table}

\begin{table}[!ht]
    \centering
    \caption{Accuracy of action prediction}
    \scalebox{1.0} {
    \begin{tabular}{cccc}
        \toprule
        \multirow{2.5}{*}{$Method$} & \multicolumn{3}{c}{$Cosine simularity$} \\ \cmidrule(lr){2-4}
        & {All} & {wo/conv} & {wo/conv-hist} \\ \midrule
         LLM-llama  & 0.220 & 0.201 & 0.176 \\ \midrule
         VLM-llama ({$d>1m$})  & 0.208 & 0.199 & 0.176 \\ \midrule
         VLM-llama ({$d>2m$})  & 0.207 & 0.198 & 0.187 \\ \midrule
         VLM-llama ({$d>3m$})  & 0.203 & 0.200 & 0.185 \\ \midrule
         Ours ({$d>1m$})\\(LLM-llama+Trajectory) & 0.218 & 0.210 & 0.207 \\ \midrule
         Ours ({$d>2m$})\\(LLM-llama+Trajectory) & 0.222 & 0.214 & 0.210 \\ \midrule
         Ours ({$d>3m$})\\(LLM-llama+Trajectory) & \textbf{0.227} & \textbf{0.218} & \textbf{0.213} \\ \midrule \midrule
         LLM-gpt  & 0.335 & 0.278 & 0.236 \\ \midrule
         VLM-gpt ({$d>1m$})  & 0.368 & 0.313 & 0.256 \\ \midrule
         VLM-gpt ({$d>2m$})  & 0.370 & 0.311 & 0.254 \\ \midrule
         VLM-gpt ({$d>3m$})  & 0.370 & 0.305 & 0.243 \\ \midrule
         Ours ({$d>1m$})\\(LLM-gpt+Trajectory) & 0.387 & 0.353 & 0.310 \\ \midrule
         Ours ({$d>2m$})\\(LLM-gpt+Trajectory) & 0.397 & 0.365 & 0.322 \\ \midrule
         Ours ({$d>3m$})\\(LLM-gpt+Trajectory) & \textbf{0.409} & \textbf{0.379} & \textbf{0.331} \\
        \bottomrule
    \end{tabular}
    }
    \label{tab:accuracy_act}
\end{table}

\subsubsection{Qualitative result}
\begin{figure}
  \includegraphics[width=\columnwidth]{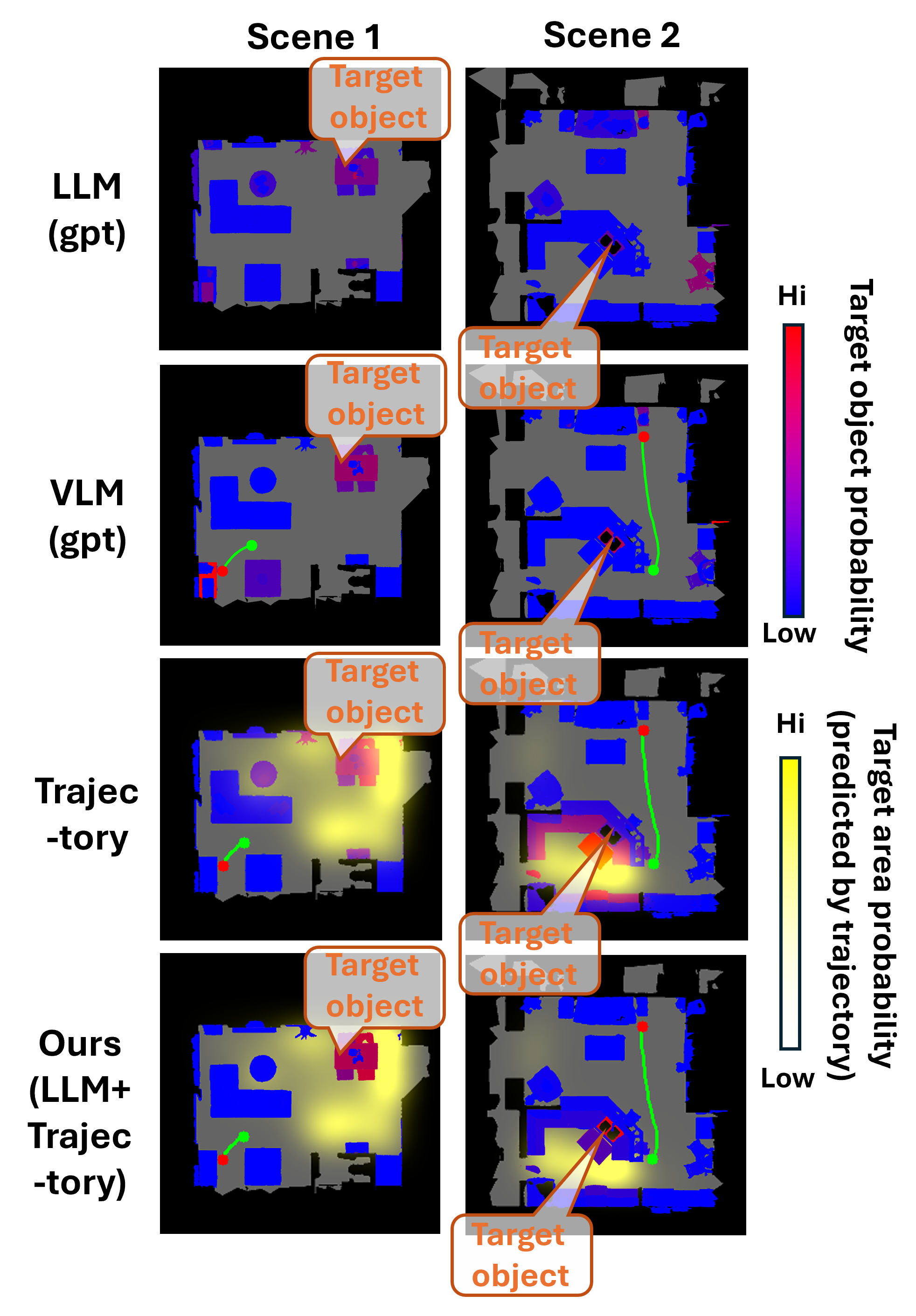}
  \caption{Target object prediction: LLM and VLM identify several object candidates, but their predictions include errors due to limited physically-aware capabilities. In contrast, while the standalone trajectory-based method roughly predicts target object areas, it lacks precision. Our method narrows down the target object more effectively by leveraging both semantic and physical cues.}
  \label{fig:Qualitative_1}
\end{figure}

\begin{figure}
  \includegraphics[width=\columnwidth]{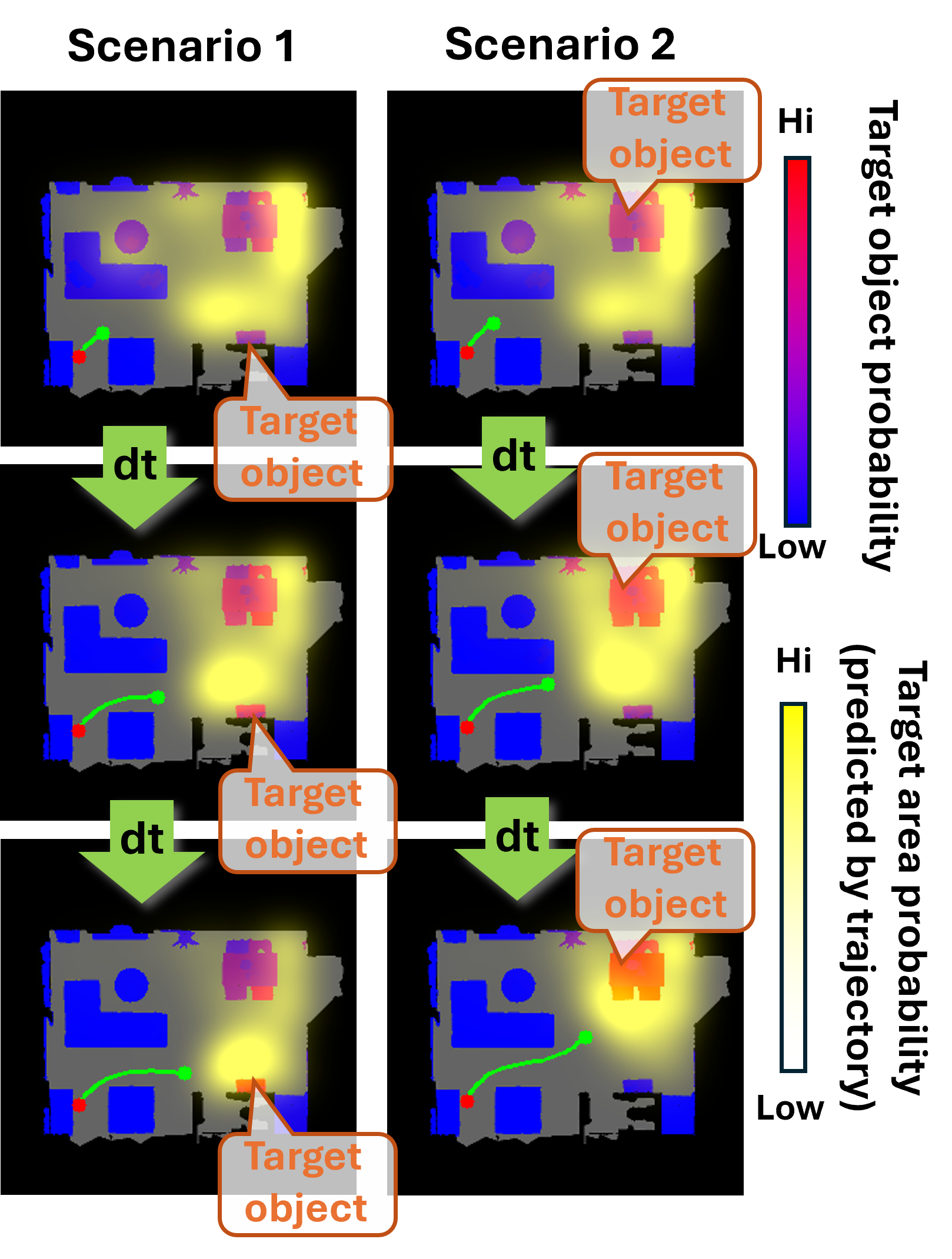}
  \caption{Behavior of trajectory-based target object prediction over time: In the top row, the target area probability distributions (highlighted in yellow) are widely spread across both scenarios. As the trajectory progresses, the distributions in the middle and bottom rows become increasingly concentrated around the ground truth target object areas.}
  \label{fig:Qualitative_2}
\end{figure}

In this section, we first present visual comparisons between our method and baseline approaches in sample scenes (Fig.\ref{fig:Qualitative_1}), and then detail how target object prediction evolves as the trajectory progresses (Fig.\ref{fig:Qualitative_2}).

Fig.\ref{fig:Qualitative_1} displays the target object prediction results using LLM-gpt, VLM-gpt, standalone trajectory-based method (Trajectory), and our proposed method (LLM-gpt+Trajectory). The red point indicates the starting position of the trajectory, while the green line and point represent the observed trajectory and current position, respectively. The yellow distribution illustrates the predicted target area based on the observed trajectory, while the objects—color-coded from blue to red—indicate predicted target object probabilities from low to high. The figure indicates that the LLM assigns high probabilities to several objects, including the target; however, some mispredictions persist due to the inherent difficulty of inferring a person’s intentions solely from the semantic scene context. Similarly, the VLM produces some incorrect predictions because of its limited capability in physically-aware prediction. Although the exact mechanism remains unclear for the pre-trained model, we observe that the VLM likely to assign relatively high probabilities to objects near the trajectory, indicating that it does not appear to predict the future target area. On the other hand, the standalone trajectory-based target object prediction assigns high probabilities widely around the ground truth target area, yet it struggles to accurately pinpoint the target object—especially when the trajectory has just begun. By leveraging both semantic and physical cues, our method more effectively narrows down the target object compared to using either LLM or trajectory data alone.

Fig.\ref{fig:Qualitative_2} illustrates the detailed behavior of the trajectory-based target object prediction by showing two distinct trajectory patterns over time within the same room layout. Initially (upper row), the target area probability distributions (shown in yellow) are broadly spread in both scenarios. As the trajectories progress (middle and bottom rows), these distributions narrow to regions near the ground truth target object, demonstrating the successful refinement process over time.

\section{Limitation and Futurework}
\textbf{Introducing Full-Body Motion:} While this study focused on trajectories as the primary form of physical information, future work could explore full-body motion, which may provide more refined and highly available cues for action prediction. Incorporating signals such as hand, foot, head, and gaze movements could serve as valuable indicators for predicting actions even during periods without locomotion.

\textbf{Application to Larger Environments:} The experiments in this study were conducted in private home environments within a 10-meter-square area. However, applying the proposed method to larger public spaces, such as schools, offices, stores, and stations, could yield greater temporal advantages in predictions through our framework. This would expand the potential for a wider range of service applications.

\textbf{Performance enhancement:} Further improvement in prediction accuracy is expected by independently enhancing both the LLM and trajectory prediction components. The refinement of each component would offer greater benefits through the integration of the two modalities.

\textbf{Application to Service Development:} A promising application of the proposed method lies in the development of intelligent systems that leverage LLMs to deliver context-aware and adaptive services tailored to predicted human actions. By anticipating user behavior, these systems could provide more timely and relevant responses, significantly enhancing user experience across various domains.

\section{Conclusion}

We leverage LLM to predict human actions incorporating diverse scene contexts in home environments. To improve the robustness against insufficient scene observations, we propose a multimodal prediction framework that combines LLM-based action prediction with physical constraints derived from human trajectories. The key idea is to integrate two different perspectives—physical and semantic factors—through an object-based action prediction framework, which compensates for the limitations of each perspective and effectively refine the prediction. The experiments show that our method markedly enhances prediction performance compared with LLM and VLM, especially in scenarios where the LLM has limited access to scene information. This improvement underscores the complementary roles of linguistic knowledge and physical constraints in comprehending and anticipating human behavior.

\appendix

\subsection{Implementation details}
\subsubsection{LLM-based prediction}
\label{appendix_llm-based_prediction}
\begin{figure}
  \includegraphics[width=\columnwidth]{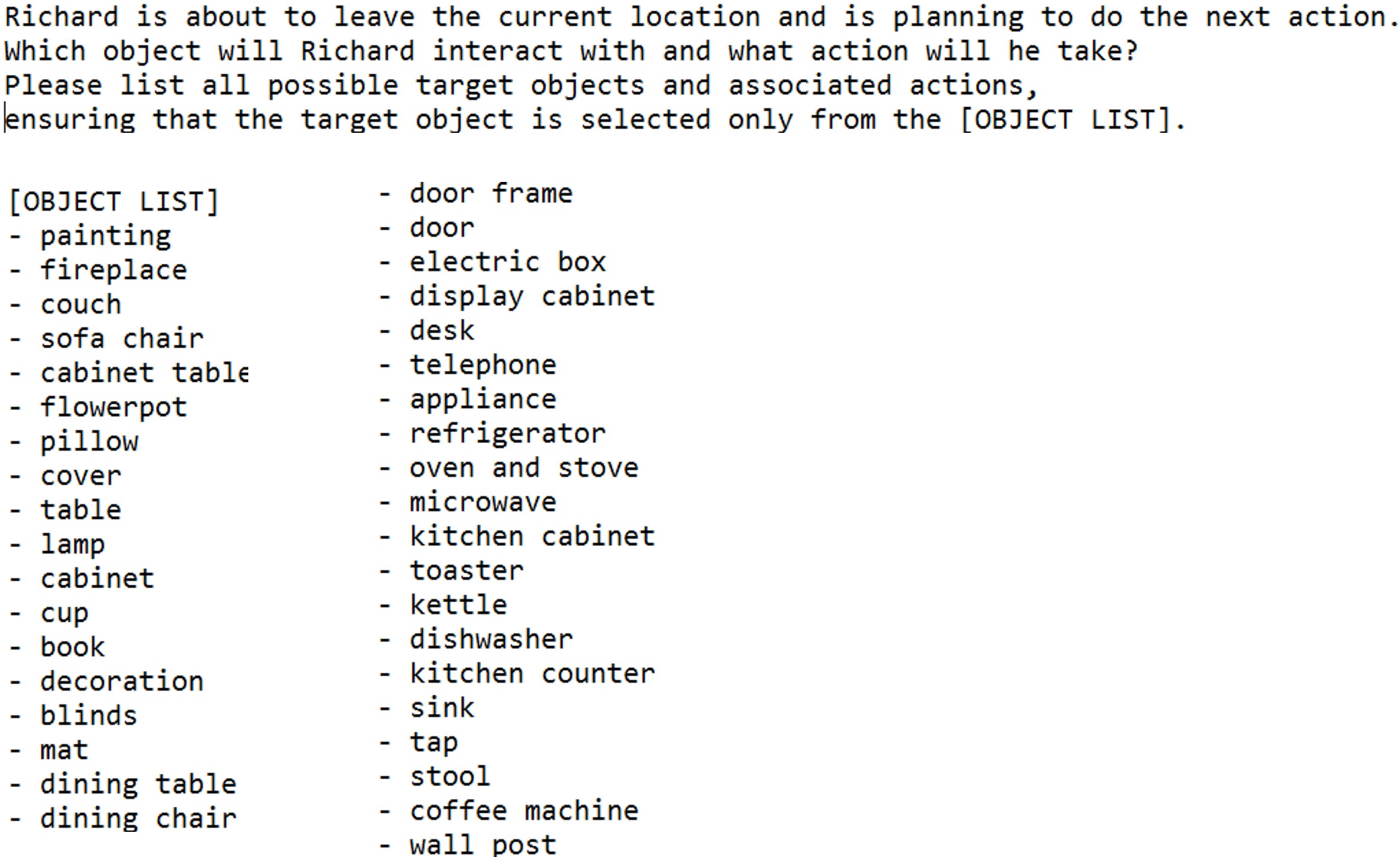}
  \caption{Order prompt to generate free-form list of potential target objects and their associated actions.}
  \label{fig:appendix_LLM_tobjpred_prompt0}
\end{figure}

\begin{figure}
  \includegraphics[width=\columnwidth]{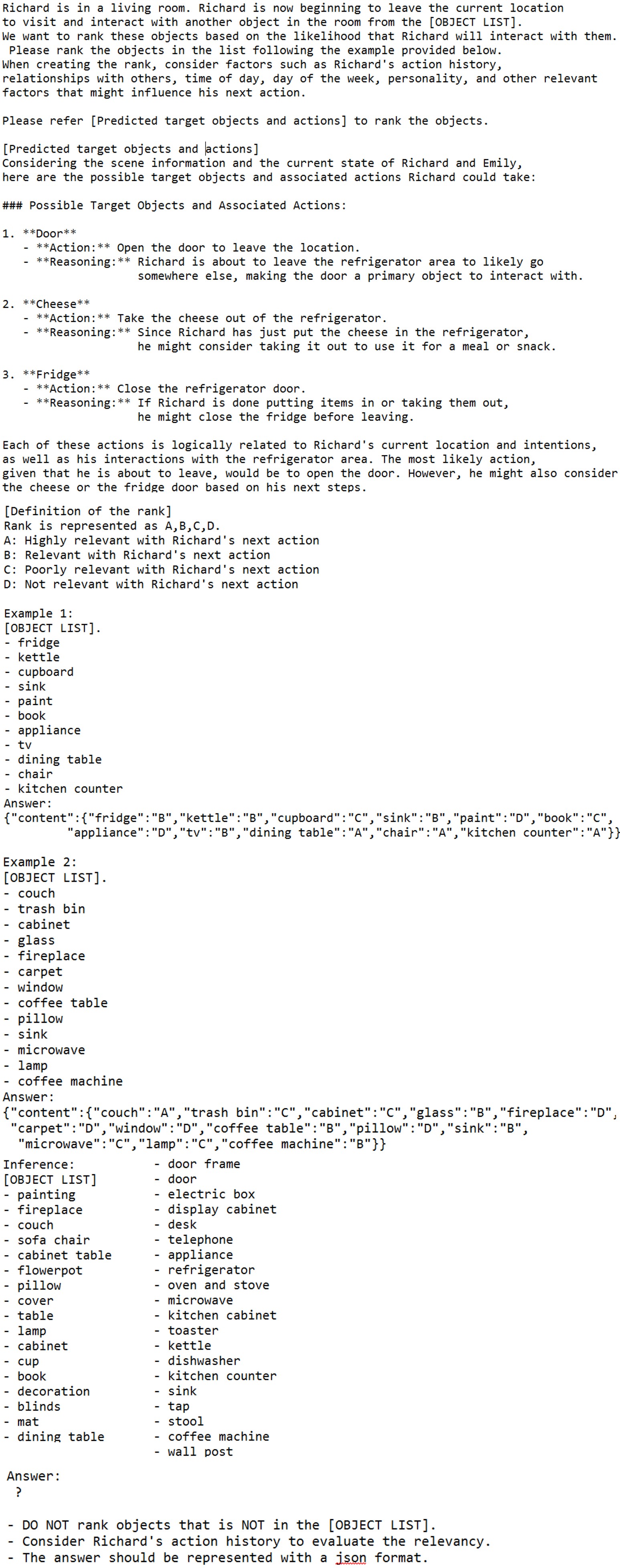}
  \caption{Order prompt to generate ranks of objects in the environment.}
  \label{fig:appendix_LLM_tobjpred_prompt1}
\end{figure}

Here, we present the specific sequence of prompts input into the LLM. Fig.\ref{fig:appendix_LLM_tobjpred_prompt0} and Fig.\ref{fig:appendix_LLM_tobjpred_prompt1} illustrate the prompts used for LLM-based target object prediction, corresponding to the first and second prompts provided to the LLM as shown in Fig.\ref{fig:LLM pipeline}, respectively. Note that [predicted target objects and actions] defined in the Fig.\ref{fig:appendix_LLM_tobjpred_prompt1} is derived from the results provided by the order prompt shown in Fig.\ref{fig:appendix_LLM_tobjpred_prompt0}.

\begin{figure}
  \includegraphics[width=\columnwidth]{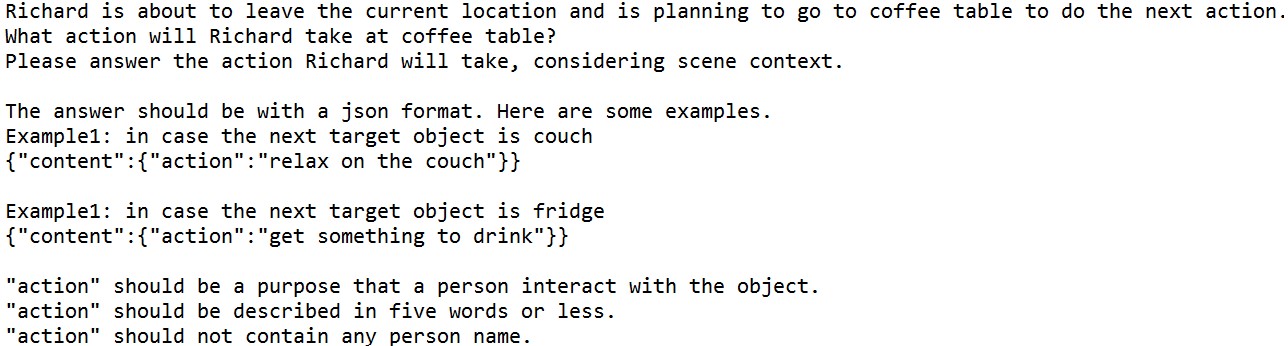}
  \caption{Order prompt to predict action based on the predicted target object.}
  \label{fig:appendix_LLM_actpred_prompt}
\end{figure}

Fig.\ref{fig:appendix_LLM_actpred_prompt} shows the order prompt used to predict an action given a target object. It is designed to generate a concise description of the action expected to occur at the specified target object.

\begin{figure}
  \includegraphics[width=\columnwidth]{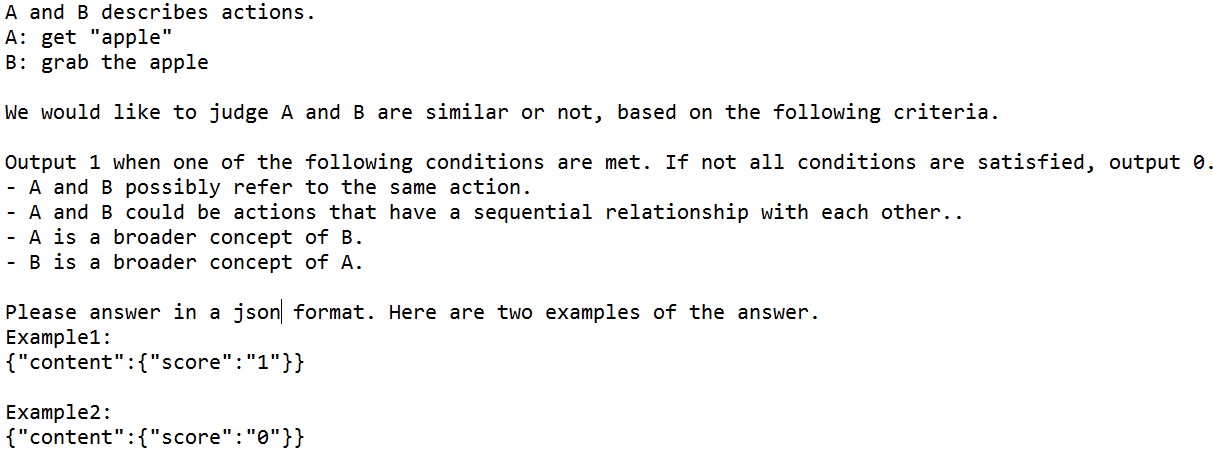}
  \caption{Order prompt used to evaluate the correctness of action prediction outcomes.}
  \label{fig:appendix_LLM_acteval_prompt}
\end{figure}

Additionally, Fig.\ref{fig:appendix_LLM_acteval_prompt} presents the order prompt used for scoring the predicted action during the evaluation stage. It outputs 1 if the predicted action has a certain similarity to the groundtruth action, and 0 otherwise.

\subsubsection{Trajectory-based prediction}
\label{appendix_trajectory-based_prediction}

\textbf{Training data}
We employed LocoVR that includes over 7000 trajectories in 131 indoor environments to train the model. We split it into training (85\%) and validation sets (15\%). 

\textbf{Model and parameters}
We employed five layer U-Net model to predict the goal area from the past trajectory. Parameters for the model are shown as below.

In the U-Net models, time-series trajectory data is represented in a multi-channel image format. Specifically, the 2D coordinates of a trajectory point are plotted on a blank 256x256 pixel image using a Gaussian distribution, with time-series data encoded across multiple channels. Similarly, the goal position is encoded as an image and concatenated with the multi-channel trajectory image as input to the model.

We employed the Adam optimizer \cite{c28} to train the model, with a learning rate of 5e-5 and a batch size of 16. Each model is trained for up to 100 epochs on a single NVIDIA RTX 4080 graphics card with 8 GB of memory.

\begin{itemize}
    \item Input: (181$\times$H$\times$W)
    \begin{itemize}
        \item Past trajectory of $p_1$ for 90 epochs (90$\times$H$\times$W)
        \item Past heading directions of $p_1$ for 90 epochs (90$\times$H$\times$W)
        \item Binary scene map (1$\times$H$\times$W)
    \end{itemize}

    \item Output: (1$\times$H$\times$W)
    \begin{itemize}
        \item $p_1$'s goal position (1$\times$H$\times$W)
    \end{itemize}

    \item Groundtruth: (1$\times$H$\times$W)
    \begin{itemize}
        \item $p_1$'s goal position (1$\times$H$\times$W)
    \end{itemize}

    \item Loss: BCELoss between the output and ground-truth

    \item U-Net channels:
    \begin{itemize}
        \item encoder: 256, 256, 512, 512, 512
        \item decoder: 512, 512, 512, 256, 256
    \end{itemize}

    \item Calculation time for training: 30-35 hours on LocoVR
\end{itemize}

\subsection{Evaluation data}
\label{appendix_data}
\subsubsection{Scenes}
\begin{figure}
  \includegraphics[width=\columnwidth]{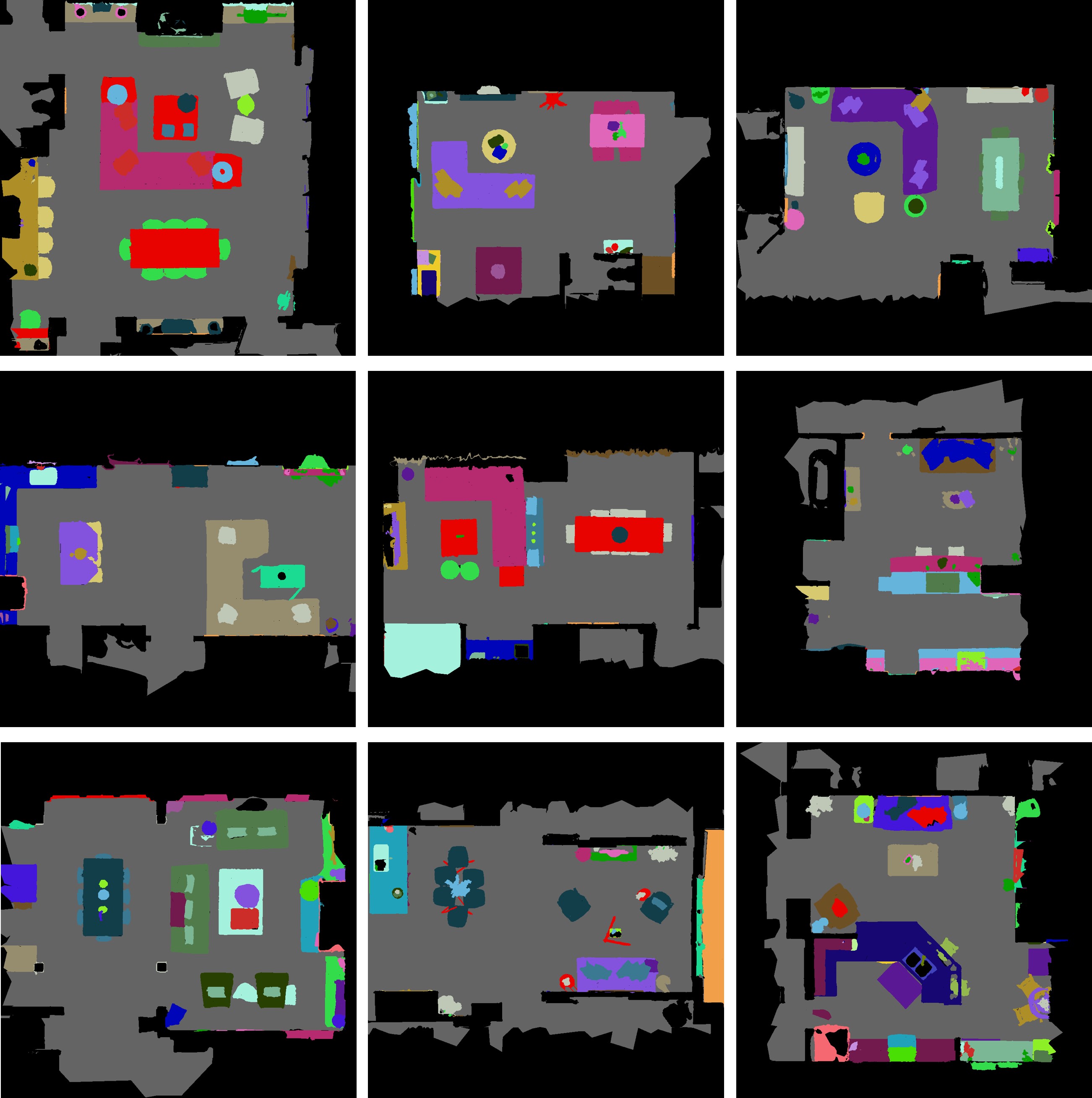}
  \caption{Scene maps and object labels}
  \label{fig:appendix_scene_maps}
\end{figure}

We converted the 3D models from HM3DSem \cite{c27} into two-dimensional scene maps, incorporating semantic labels and corresponding object pixelations. Fig.\ref{fig:appendix_scene_maps} illustrates these scene maps along with their associated object labels. Each scene contains approximately 30 to 40 distinct objects, with each object represented by a unique color.

\subsubsection{Trajectory synthesis}
\label{appendix_trajectory_synthesis}
To evaluate the performance of our multimodal human action prediction model, we utilized evaluation data that includes human trajectory information. Given that our method predicts target areas based on observed trajectories, it is crucial that the trajectory data be realistic in terms of both spatial configuration and velocity profile. To ensure this, we employed a machine learning-based approach for trajectory generation. Using the LocoVR dataset, which provides comprehensive data on start positions, target positions, scene maps, and trajectories, the model was trained to learn the relationships between these elements. Specifically, the model takes start positions, target positions, and scene maps as inputs and predicts the corresponding trajectories. By incorporating both trajectory shape and velocity variations into the loss function, the model is able to generate realistic position-velocity sequences.

\end{document}